\documentclass[pra,aps,nofootinbib,notitlepage,showpacs]{revtex4-1}
\input{Qcircuit}
\usepackage{amsfonts}
\usepackage{amsmath,amssymb}
\usepackage{physics}
\usepackage{tikz}
\usepackage{bm}
\usepackage{xcolor}
\usepackage{hyperref}
\hypersetup{
     colorlinks   = true,
     citecolor    = blue
}

\def\be{\begin{equation}}
\def\ee{\end{equation}}
\def\bea{\begin{eqnarray}}
\def\eea{\end{eqnarray}}

\begin{document}
\title{Quantum Imaginary Time Evolution Algorithm for Quantum Field Theories with Continuous Variables}
{\footnote{This manuscript has been authored by UT-Battelle, LLC, under Contract No. DE-AC0500OR22725 with the U.S. Department of Energy. The United States Government retains and the publisher, by accepting the article for publication, acknowledges that the United States Government retains a non-exclusive, paid-up, irrevocable, world-wide license to publish or reproduce the published form of this manuscript, or allow others to do so, for the United States Government purposes. The Department of Energy will provide public access to these results of federally sponsored research in accordance with the DOE Public Access Plan.}}
\author{K\"ubra Yeter-Aydeniz }
\email{yeteraydenik@ornl.gov}
\affiliation{Physics Division, Oak Ridge National Laboratory,
  Oak Ridge, TN 37831, USA}
  \affiliation{Computational Sciences and Engineering Division, Oak Ridge National Laboratory,
  Oak Ridge, TN 37831, USA}
\author{Eleftherios Moschandreou }
\email{emoschan@vols.utk.edu}
\affiliation{Department of Physics and Astronomy,  The University of Tennessee, Knoxville, TN 37996-1200, USA}
\author{George Siopsis }
\email{siopsis@tennessee.edu}
\affiliation{Department of Physics and Astronomy,  The University of Tennessee, Knoxville, TN 37996-1200, USA}
\date{\today}

\begin{abstract}
    We calculate the energy levels and corresponding  eigenstates of an interacting scalar quantum field theory on a lattice using a continuous-variable version of the quantum imaginary time evolution algorithm. Only a single qumode is needed for the simulation of the field at each point on the lattice. Our quantum algorithm avoids the use of non-Gaussian quantum gates and relies, instead, on detectors projecting onto eigenstates of the photon-number operator. Using Xanadu's Strawberry Fields simulator, we obtain results on energy levels that are in very good agreement with results from exact calculations. We propose an experimental setup that can be realized with existing technology.
\end{abstract}
\maketitle
\section{Introduction}
\label{sec:intro}







Quantum fields are fundamental constituents of the physical world describing quantum many-body systems of matter at all energy scales, as well as electromagnetic and gravitational radiation.  Quantum-field engineering has enabled unprecedented measurement sensitivities, epitomized by the use of squeezed light to lower the noise floor of the Laser Interferometer Gravitational-wave Observatory (LIGO) below the shot noise limit~\cite{Aasi2013}.

The encoding of quantum information in continuous-variable (CV) quantum fields, a.k.a.\  qumodes (in lieu of discrete-variable (DV) qubits), has enabled multipartite entanglement over millions of qumodes.  This scale, unparalleled in any qubit architecture, defines new horizons and paradigms for quantum computing, quantum communication, and quantum sensing.  Nanophotonic integrated devices based on qumodes have the potential to define future quantum technology by surpassing the performance of qubit-based Noisy Intermediate-Scale Quantum (NISQ)~\cite{Preskill2018} computing devices.

A natural implementation of qumodes uses quantum light, which also lends itself to sensing~\cite{Xia2021, Casacio2021, Pooser2020,Lawrie2019} and communication~\cite{Xu2020, Pirandola2020}.  The coming of age of low-loss, high-nonlinearity integrated optics paves the way for implementing large-scale, fault-tolerant quantum computing and communication devices on chip, at room temperature, and within a few years~\cite{Sciarrino2020}.

Over the last 20 years, since the introduction of the first quantum algorithm by Deutsch and Jozsa~\cite{Deutsch-Jozsa1992}, even though a tremendous amount of DV quantum algorithms has been proposed capable of solving various problems more efficiently than their classical counterparts, there has been far less activity in the development of CV quantum computing. Examples of quantum algorithms generalized to CV substrates include the Deutsch-Jozsa~\cite{Adcock2009} and Grover's search~\cite{Pati2000} algorithms, and more recently the quantum approximate optimization algorithm (QAOA)~\cite{Verdon2019} and a CV quantum algorithm for solving linear partial differential equations~\cite{Arrazola2019}. 

There has also been an effort to extend quantum machine learning algorithms to a CV substrate~\cite{Lau2017} followed by the singular value decomposition of nonsparse, low rank matrices in~\cite{Das2018}, and topological data analysis~\cite{Siopsis2019}. CV quantum neural networks were introduced in~\cite{Killoran2019} and used in applications such as fraud detection with a CV classifier, a hybrid classical-quantum auto encoder. 

Considering the recent experimental breakthroughs in CV photonic quantum computing \cite{Zhong2020} and the development of a programmable photonic quantum computer chip by the Xanadu team \cite{Arrazola2021}, it is important to explore CV quantum algorithms applied to realistic experimental setups.

Here, we propose a CV quantum imaginary-time evolution (QITE) algorithm which can be used to calculate the energy levels and corresponding energy eigenstates of an interacting scalar quantum field theory (QFT) on a lattice. A QFT for a massive self-interacting scalar field $\phi$ with a $\phi^4$ interaction was studied in \cite{Jordan2012} where it was shown that quantum algorithms for scattering amplitudes provided an exponential speed up over known classical algorithms. 
QFTs were further studied in~\cite{YeterAydeniz2019QFT, Barata2021, Davoudi2021} with DV quantum algorithms and in~\cite{Marshall2015, YeterAydeniz2018QED} on a CV substrate for a $\phi^4$ scalar QFT as well as scalar quantum electrodynamics (QED). There is a growing interest in simulating lattice gauge theories using quantum computers as well~\cite{Banuls2020}.

There have been a few alternative approaches. In Ref.~\cite{Liu2020}, quantum computation of two-dimensional quantum chromodynamics (QCD) was discussed without introducing a lattice.
A different perspective was offered in~\cite{Milsted2021} where it was proposed that the noise of NISQ hardware can be avoided by simulating the dynamics of QFTs using classical simulators and tools such as Matrix Product States (MPS). 
 
CV quantum algorithms for QFTs have important advantages over their DV counterparts. Their implementation generally relies on optical elements which operate at room temperature. Moreover, while DV algorithms need a whole register of qubits, only a single qumode is needed at each point on the lattice to simulate the field at that point. However, CV quantum algorithms that have been proposed so far involve non-Gaussian quantum gates which are hard to implement with existing technology \cite{bibCV1,bibCV2,bibCV3,bibCV4}. Here, we avoid the introduction of non-Gaussian gates by employing a CV adaptation of the QITE algorithm making use of detectors projecting onto eigenstates of the photon number operator which is readily available technology.

Imaginary-time evolution has been extensively used in studying quantum many-body systems as a useful tool for various tasks, such as the calculation of the ground state energy and the creation of finite-temperature states. Evolution in imaginary time $\tau$ is implemented with the non-unitary operator $\mathcal{U} (\tau)=e^{-\tau H}$ where $H$ is the Hamiltonian of the system of interest. Starting with an initial state that has non-zero overlap with the ground state of the system, the evolved state convergences to the ground state in the limit $\tau \to \infty$. Excited states can also be reached with an appropriate choice of initial state (one that is orthogonal to the ground state). The calculation of the energy spectrum of a many-body quantum system is a daunting yet important task as it provides important information about the system. As the number of particles increases, the calculation becomes exponentially harder. Then it becomes imperative to employ the quantum version of the imaginary-time evolution algorithm as it outperforms its classical counterpart. 

Simulating the imaginary-time evolution on a quantum computer is not straightforward because $\mathcal{U}(\tau)$ is a non-unitary operator. Various approaches to the implementation of $\mathcal{U} (\tau)$ have been proposed. A variational version of QITE was proposed in~\cite{McArdle2019} in which the Wick-rotated Schr\"odinger equation was solved for parameterized states and a classical optimization loop was used to estimate the parameters for the ground state of molecular Hydrogen and the LiH molecule. Although this method offers shallow quantum circuits, the classical optimization of the parameters becomes prohibitive as the number of particles in the system grows. Motta \textit{et al.}\ \cite{Motta2019} proposed a QITE algorithm that did not require classical optimization or an ancilla qubit. They expressed the Hamiltonian in terms of local terms and used Trotterization to implement $\mathcal{U} (\tau)$. Then the non-unitary evolution operator for a small imaginary-time interval was approximated by a unitary operator which was expressed in terms of Pauli spin operators with coefficients calculated from measurements on quantum hardware. The problem with this method is that the number of measurements grows exponentially with the system size and the quantum circuit becomes longer with each QITE step making it hard to implement on NISQ devices. Various attempts to bypass this problem have been made~\cite{YeterAydeniz2019, YeterAydeniz2020, Sun2020, Nishi2021, Shirakawa2021}. 

We implement the QITE algorithm by approximating the non-unitary evolution operator with a Gaussian operator, $\mathcal{U} (\tau) \approx e^{-\tau \mathcal{A}}$, where $\mathcal{A}$ is a Hermitian operator and a function of one of the two quadratures, $q$. After Trotterization, the \emph{Ansatz} $\mathcal{A}$ is determined at each step in a manner similar to the DV approach in~\cite{Motta2019}. The non-unitary Gaussian operator $e^{-\tau \mathcal{A}}$ is realized with the aid of ancilla qumodes. Only a single ancilla is needed at each step regardless of the size of the system. As the number of steps increases, the length of the quantum circuit does not increase indefinitely, unlike in the DV case, because the contributions to the Gaussian operator $\mathcal{A}$ at each QITE step commute with each other. Thus the quantum circuit only involves a finite number of parameters which are determined by quantum measurements at each QITE step. 
Thus, we avoid the use of non-Gaussian elements and rely on quantum measurements that involve detectors projecting onto photon-number eigenstates. We explain how our quantum algorithm can be realized with existing technology.

Our discussion is organized as follows. In Section~\ref{sec:Model}, we review the discretization of the massive $\phi^4$ self-interacting scalar QFT. In Section~\ref{sec:QITE} we discuss the details of our CV quantum algorithm. 
In Section~\ref{sec:sim_res}, we discuss our results using Xanadu's Strawberry Fields CV photonic quantum simulator~\cite{Killoran2019SF, Bromley2020} and show that they are in agreement with exact results. We also outline an experimental realization of the quantum circuit with existing technology. We present our conclusions and outlook in Section~\ref{sec:conclusion}.

\section{The Model}
\label{sec:Model}
For definiteness, we concentrate on the simplest QFT describing a massive self-interacting scalar field in one spatial dimension with a quartic interaction. Our results can be generalized to more complicated QFTs, including gauge theories that describe elementary particle interactions.

To study the QFT on a quantum computer, we discretize the system in space with coordinate $x=0,1,\dots,L-1$ where $L$ is the length of the spatial dimension in units in which the lattice spacing is $a=1$. The Hamiltonian for a massive scalar field $\phi(x)$ with a quartic interaction term is 
\be
H=\sum_{x=0}^{L-1}\left[ \frac{1}{2}\pi^2(x)+ \frac{1}{2}\left[\nabla \phi(x)\right]^2+\frac{m_0^2}{2} \phi^2(x)+\frac{\lambda}{4!}\phi^4(x)\right] ~,
\ee
where $m_0$ is the bare mass of the scalar field, $\lambda$ is the interaction strength, and $\pi(x)$ is the conjugate momentum obeying the commutation relations
\be
\left[\phi(x),\pi(x')\right]=i\delta_{xx'}~.
\ee
The scalar field and its conjugate momentum can be expressed in terms of creation and annihilation operators obeying commutation relations
$\left[a(k),a^\dagger(k')\right]=\delta_{kk'}$ as
\be
\phi(x)=\frac{1}{\sqrt{L}}\sum_{k=0}^{L-1}\frac{1}{\sqrt{2\omega(k)}}\left[a^\dagger(k)e^{-2\pi ikx/L}+\text{h.c.} \right]~,
\ee
\be
\pi(x)=\frac{i}{\sqrt{L}}\sum_{k=0}^{L-1} \sqrt{\frac{\omega(k)}{2}}\left[a^\dagger(k)e^{-2\pi ikx/L}- \text{h.c.} \right]
\ee
where
\be \omega (k) = \sqrt{m^2 + 4\sin^2 \frac{\pi k}{L}}~. \ee
The mass parameter $m$ is arbitrary as long as $m^2 > 0$. One may choose $m = m_0$, but this is often not possible in physically interesting cases in which $m_0^2 < 0$.
If $m$ is chosen as the physical mass parameter, $m = m_{\text{phys}}$ (notice that in an interacting system $m_{\text{phys}} \ne m_0$, due to quantum effects), then in the continuum limit, $\omega(k)$ is the energy of a relativistic particle of momentum $\frac{2\pi k}{L}$. No physical quantities depend on the choice of $m$, although spurious dependencies enter if one simulates the system with qubits due to the truncation of the Hilbert space. This is not an issue with CV quantum algorithms.

Having chosen $m$, we define the mass counter term $\delta m$ by $m^2=m_0^2+\delta m$. The Hamiltonian splits into a non-interacting ($H_0$) and interacting ($H_I$) piece,
\be
H=H_0+H_I
\ee
where
\be
H_0 = \frac{1}{2}\sum_{x=0}^{L-1}\left[\pi^2(x)+\left[\nabla \phi(x)\right]^2+m^2 \phi^2(x)\right]~, \ \ \
H_I = \sum_{x=0}^{L-1} \left[ -\frac{\delta m}{2} \phi^2(x)+ g\phi^4(x) \right]~.
\ee
The non-interacting Hamiltonian is diagonal in the momentum representation,
\be\label{eq:9}
H_0=\sum_k  \omega(k) \left(a^\dagger(k) a(k)+\frac{1}{2}\right)
\ee
whereas the interaction part is diagonal in the position representation.

For the CV quantum calculation, it is convenient to work with the quadratures
\be q(k) = \frac{1}{\sqrt{2}} \left[ a^\dagger (k) + a (k) \right] \ , \ \ 
p(k) = \frac{i}{\sqrt{2}}  \left[ a^\dagger (k) - a (k) \right]
\ee 
in terms of which the non-interacting
Hamiltonian reads
\be H_0 =  \sum_k \frac{\omega(k)}{2} \left[ p^2(k) + q^2 (k)  \right]~. \ee
Its ground state can be written as
\be\label{eq:Omega0}
|\Omega_0\rangle=|0\rangle_0 \otimes |0\rangle_1 \otimes \dots \otimes |0\rangle_{L-1}~, 
\ \ 
\langle \bm{q}|\Omega_0 \rangle = \frac{1}{\pi^{L/4}} e^{-\bm{q}^2/2}
\ee
where $\bm{q} = (q(0), \dots , q(L-1))$. It has zero energy, by design.

Alternatively, we can work with the quadratures in the position representation, $\phi(x), \pi(x)$.
The two sets of quadratures are related to each other by a Bogoliubov transformation that can be implemented with beam splitters and squeezers,
\bea
\phi(x) &=& \frac{1}{\sqrt{L}}\sum_{k=0}^{L-1}\frac{1}{\sqrt{\omega(k)}}\left[ q(k) \cos \frac{2\pi kx}{L} - p(k) \sin \frac{2\pi kx}{L} \right]~,
\nonumber\\ \pi(x) &=& \frac{1}{\sqrt{L}}\sum_{k=0}^{L-1} \sqrt{\omega(k)} \left[ p(k) \cos\frac{2\pi kx}{L} + q(k) \sin\frac{2\pi kx}{L} \right]~.
\eea
\section{CV Quantum Algorithm}
\label{sec:QITE}

In this section, we introduce our CV quantum algorithm for the calculation of energy levels and corresponding eigenstates of the scalar QFT.

\subsection{Ground state}
We start by introducing a CV version of the QITE algorithm. We choose the initial state \eqref{eq:Omega0},
\be
|\Psi[0]\rangle=|\Omega_0\rangle
\ee
and obtain the estimate of the ground-state energy of $H$,
\be E[0] = \langle \Psi[0]|H|\Psi[0]\rangle~. \ee
The imaginary-time evolution operator can be factored into small (Trotter) steps, 
\be
e^{-\tau H}=( e^{-\Delta \tau H})^n~,
\ee
where $\tau=n\Delta \tau$. We wish to approximate each step that involves non-Gaussian operations with an operator that can be efficiently computed with a CV quantum algorithm. 
To describe the iterative process, suppose that after $s-1$ steps, we arrive at the state $|\Psi[s-1]\rangle$. In the $s$th step, we evolve this state in small imaginary time, $\Delta\tau$.
The evolved state is
\be
|\Psi_s(\Delta \tau)\rangle \equiv \frac{e^{-\Delta \tau H } |\Psi[s-1]\rangle}{\| e^{-\Delta \tau H} |\Psi[s-1]\rangle\|}~. \label{Psis}
\ee
To approximate this non-Gaussian state, we introduce the $s$th step Gaussian \emph{Ansatz},
\be |\Psi[s]\rangle  = \frac{e^{ - \Delta\tau\sum_k \gamma_s(k) q^2(k)/2 } |\Psi[s-1]\rangle }{\| e^{ - \Delta\tau \sum_k \gamma_s(k) q^2(k) /2 } |\Psi [s-1]\rangle \| }  \label{eq:Psis} \ee
and choose the parameters $\gamma_s (k)$
($k=0,\dots, L-1$) that minimize the distance $\| |\Psi[s]\rangle-|\Psi_s(\Delta \tau)\rangle \|$ at first order in $\Delta \tau$. We obtain
\be \| |\Psi[s]\rangle - |\Psi_s(\Delta\tau)\rangle \|^2 \approx (\Delta\tau)^2 \sum_k \mathcal{X}_s (k) + \text{const.} \ee
where
\be 
\mathcal{X}_s (k)  = \frac{1}{4} ( \langle q^4(k) \rangle - \langle q^2(k) \rangle^2 )\gamma_s^2(k)    -  \left(  \langle q^2(k) H \rangle-  \langle q^2(k) \rangle \langle H\rangle \right)  \gamma_s(k) 
	\ee 
with all expectation values calculated with respect to $|\Psi[s-1]\rangle$.
The distance is minimized for the choice of paramaters
\be \gamma_s(k) = 2\frac{\langle q^2(k) H  \rangle - \langle q^2(k) \rangle \langle H\rangle}{ \langle q^4(k) \rangle -\langle q^2(k) \rangle^2 }~. \label{gamma_sq}\ee
These parameters can be calculated at each step from expectation values obtained from CV quantum hardware. Thus, the state starting with \eqref{eq:Omega0} after $s$ QITE steps \eqref{eq:Psis} is a squeezed state~\cite{Braunstein2004},
\be \langle \bm{q} | \Psi [s] \rangle \propto \otimes\prod_{k=0}^{L-1} e^{-\frac{q^2(k)}{2}\sigma_s^2(k)} \ , \ \ \sigma_s^2(k) = 1 + \Delta\tau \sum_{s'=1}^s \gamma_{s'} (k)~, \ee
and can be realized on a CV substrate with single-mode squeezers of respective squeezing parameters 
$r_s(k)= \log\sigma_s(k)$.


Although the above method leads to fast convergence in the QFT considered here, for completeness we present an alternate method that can be applied to a general choice of initial state, since more complex initial states may be needed for convergence in more complicated systems. Starting from an unspecified initial state $|\Psi[0]\rangle$, we introduce an ancillary qumode in the vacuum state $|0\rangle_{\text{anc}}$.
Then we apply the controlled-addition ($CX$) gate (a Gaussian), 
\be\label{eq:22} CX(\Gamma_s(k)) = e^{i\Gamma_s(k) p_{\text{anc}} \otimes q(k)} \ee 
where $\Gamma_s(k) \in \mathbb{R}$ is a parameter to be determined.
$CX(\Gamma)$ is a two-mode gate that can be decomposed into single-mode squeezers ($S$) and beam splitters ($B$) as
\be\label{eq:23} CX(\Gamma)=B\left(\frac{\pi}{2}+\theta,0\right)\left(S(r,0) \otimes S(-r,0)\right)B(\theta,0)
\ee
where 
$ \sin 2\theta=-\frac{1}{\cosh r}$, $\sinh r = - \frac{\Gamma}{2}$, and $r$ is a
squeezing parameter.

After the implementation of the $CX$ gate, the state becomes entangled:
\be
\label{eq:U2app} CX(\Gamma_s(k))  |\Psi[0]\rangle |0\rangle_{\text{anc}} 
= \int d^L \bm{q} \Psi[0] (\bm{q}) 
 |\bm{q}\rangle | \Gamma_s(k) q(k)  \rangle_{\text{anc}}~. 
\ee 
We then measure the ancilla qumode with a photon detector. If the detector detects no photon, the state collapses to (unnormalized)
\be 
 {}_{{\text{anc}}}\langle 0| CX(\Gamma_s(k)) |\Psi[0]\rangle |0\rangle_{\text{anc}} 
    \propto   e^{-\Gamma_s^2(k) q^2(k)/2 } |\Psi[0] \rangle ~.\label{eq27}
 \ee 
 We repeat the above process with different parameters for all modes and arrive at the (unnormalized) state
 \be\label{eq:26} e^{-\sum_k \Gamma_s^2(k) q^2(k)/2 } |\Psi[0] \rangle ~.\ee 
 This matches the desired state $|\Psi[s]\rangle$ with the choice of parameters determining the $CX$ gate,
 \be\label{eq:27} \Gamma_s^2 (k) = \Delta\tau \sum_{s'=1}^s \gamma_{s'} (k)~.\ee

Next, we discuss the quantum computation of the parameters $\gamma_s (k)$. They are given in terms of expectation values of even powers of the quadratures for the various modes in the state $|\Psi[s-1]\rangle$ obtained after $s-1$ QITE steps.

For $\langle q^{2n} (k) \rangle$ ($n=1,2,\dots$), we work as follows. We introduce an ancillary mode in the vacuum state, $|0\rangle_{\text{anc}}$, and apply the $CX$ gate,
\be 
    CX(\eta) |\Psi[s-1]\rangle |0\rangle_{{\text{anc}}}  = \int d^L \bm{q} \Psi[s-1] (\bm{q}) 
 |\bm{q}\rangle \left| \eta q(k) \right\rangle_{{\text{anc}}}~.
\ee
Then we measure the photon number in the ancillary mode. If the measurement outcome is $n_{{\text{anc}}} = 0$, the state is projected onto
\be 
    {}_{{\text{anc}}} \langle 0| CX(\eta) |\Psi[s-1]\rangle |0\rangle_{{\text{anc}}}   
 =e^{-\eta^2 q^2 (k)/4}|\Psi[s-1]\rangle~. \ee
The probability of this outcome is
\be\label{eq:30}
    P_{0}(k) = \left\| e^{-\eta^2 q^2 (k)/4} |\Psi[s-1]\rangle \right\|^2 =  \langle \Psi[s-1] | e^{-\eta^2 q^2 (k)/2} |\Psi[s-1]\rangle~. 
\ee
Therefore, we managed to express the expectation value $\langle e^{-\eta^2 q^2 (k)/2}\rangle$ as the probability $P_0(k)$ of a measurement outcome projecting the ancillary mode onto a photon-number eigenstate. By varying $\eta$, we obtain the expectation value of any even power $\langle q^{2n} (k) \rangle$,
\be \langle q^{2n} (k) \rangle = \left. (-2)^n \frac{d^n}{d(\eta^2)^n} P_0 (k) \right|_{\eta = 0}~. \label{eq:q2nk}\ee
By repeating this process for a second mode, we obtain all expectation values of the form $\langle q^{2n} (k) q^{2n'} (k')\rangle$, and similarly for products involving more modes. To calculate the derivatives on CV quantum hardware, a finite-differences method can be used. One may also use the parameter shift rule for CVs discussed in~\cite{Schuld2019}. In each case, the circuit is repeatedly run with different $CX$ gate parameters to obtain the gradient or higher-order derivatives of the physical quantities of interest.  

We obtain expectation values involving the $p$-quadrature by following the above procedure with $CX$ replaced by the $CZ$ gate,
\be CZ (\Gamma) = e^{i\Gamma q_{\text{anc}} \otimes q} \ee
%
These results can be used to calculate all expectation values in \eqref{gamma_sq} and therefore yield $\gamma_s(k)$. We also obtain the energy at each step converging to the ground-state of the system.


\subsection{Mass gap}

The above method may also be used to compute minimum energies of states with an odd number of excitations. There are $L$ such states corresponding to the $L$ qumodes in the system. To obtain each of these energies, we may start from the state 
\be\label{eq:33} |\Omega (k) \rangle \propto q(k)|\Omega_0\rangle \ , \ \ \langle \bm{q}|\Omega(k)\rangle \propto q(k) e^{-\bm{q}^2/2} \ , \ \ k = 0,1,\dots, L-1 \ , \ee
which has energy $\omega(k)$ in the non-interacting system. Notice that all these states are orthogonal to the ground state and form an orthonormal set ($\langle \Omega_0 |\Omega(k)\rangle = 0$, and $\langle \Omega(k)|\Omega(k') \rangle = \delta_{kk'}$). Therefore, the QITE algorithm is expected to converge to different energy levels of the system if we choose one of these states corresponding to single-particle states of momentum $\frac{2\pi k}{L}$. By momentum conservation, there can be no transitions between these states as well as between one of these states and the ground state. This is confirmed by an explicit calculation showing that transition amplitudes vanish ($\langle \Omega_0 |H|\Omega (k) \rangle = 0$, and $\langle \Omega (k) | H | \Omega(k') \rangle = 0$ if $k'\ne k$).

To calculate the mass gap, apart from the ground-state energy $E_0$, we need the first-excited-state energy $E_1$. Then the mass gap is $m_{\text{phys}} = E_1 - E_0$. To calculate $E_1$, we apply the same QITE algorithm we outlined above for the ground-state energy $E_0$, except that we choose $|\Omega(0)\rangle$ as the initial state. It is orthogonal to the ground state and has energy $\omega(0) = m$ in the non-interacting system. Thus, we initialize the QITE algorithm with the state
\be \langle \bm{q}|\Psi[0]\rangle \propto q(0) e^{-\bm{q}^2 /2} \ee
Following the same procedure as before, we adopt the \emph{Ansatz} \eqref{eq:Psis} at the $s-1$ QITE step and obtain the expression \eqref{gamma_sq} for the parameters $\gamma_s(k)$. To create the state $|\Psi[s]\rangle$, we start with the even state $|\Omega_0\rangle$ and introduce an ancilla qumode in the vacuum state. Then we apply the $CX$ gate \eqref{eq:22}, thus obtaining the entangled state \eqref{eq:U2app}. Then we measure the photon number in the ancilla qumode, but unlike in the case of the ground-state energy, for $k=0$, we project onto the single-photon state. Thus, instead of \eqref{eq:26}, we obtain the (unnormalized) state
\be\label{eq:26a} q(0) e^{-\sum_k \Gamma_s^2(k) q^2(k)/2 } |\Omega_0 \rangle ~.\ee 
with the choice of parameters given by \eqref{eq:27} for the desired state. This state approximates the first excited state of the system.

\subsection{Excited states}

The states $|\Psi[s]\rangle$ found in the course of implementing QITE can be used to extract information about the excited states of the system via the QLanczos algorithm. It should be noted that one obtains states in different sectors of the Fock space  depending on the choice of initial state, e.g., the ground state $|\Omega_0\rangle$ or one of the single-particle states $|\Omega(k)\rangle$ of the free system.

To illustrate the QLanczos algorithm, let us select two states obtained via QITE, $|\Psi[s_1]\rangle$ and $|\Psi[s_2]\rangle$, where $s_2-s_1$ is an even number. We then find the $2\times2$ Hamiltonian which is the restriction of $H$ in the subspace (Krylov space) spanned by the chosen QITE states,
\be \mathcal{H}_{ij} = \langle \Psi[s_i]|H|\Psi[s_j]\rangle \ , \ \  i,j=1,2~.\ee
We also calculate the overlap matrix
\be \mathcal{T}_{ij} = \langle \Psi[s_i]|\Psi[s_j]\rangle \ , \ \  i,j=1,2~.\ee
We obtain estimates of the energies of the ground and 2nd excited state by solving the generalized eigenvalue equation
\be \mathcal{H} \bm{x} = E \mathcal{T} \bm{x} \ee 
and estimates of the corresponding eigenstates from the corresponding eigenvectors $\bm{x}(E)$,
\be \Psi [E] = x_1(E) \Psi[s_1] + x_2(E) \Psi[s_2]~. \ee
The matrix elements $\mathcal{H}_{ij}$ and $\mathcal{T}_{ij}$ can be deduced in the course of the QITE algorithm. The state $|\Psi[s]\rangle$ is an approximation to the state 
\be |\Psi[s]\rangle \approx c_s e^{-s\Delta\tau H} |\Psi[0]\rangle \approx \frac{c_s}{c_{s-1}} e^{-\Delta\tau H} |\Psi[s-1]\rangle  \ee where $c_s$ is a normalization constant, and
\be \left( \frac{c_s}{c_{s-1}} \right)^{-2} =  \langle \Psi[s-1] |e^{-2\Delta\tau H} |\Psi[s-1]\rangle \ee
Therefore, the normalization constants can be calculated recursively, starting with $c_0=1$. We need $\langle e^{-2\Delta\tau H} \rangle\approx 1 - 2\Delta\tau \langle H\rangle$, which is found in the QITE algorithm. 

The matrix elements are obtained from QITE using
\be \mathcal{T}_{11} = \mathcal{T}_{22} =1 \ , \ \ \mathcal{T}_{12} = \mathcal{T}_{21} = \frac{c_{s_1} c_{s_2}}{c_{\bar{s}}} \ , \ \ \bar{s} = \frac{s_1+s_2}{2} \ee
and
\be \mathcal{H}_{ii} = \langle \Psi[s_i]|H|\Psi[s_i]\rangle \ , \ \ \mathcal{H}_{12} = \mathcal{T}_{12} \langle \Psi[\bar{s}] | H | \Psi[\bar{s}] \rangle~. \ee
For higher excited states, we need to consider a higher-dimensional Krylov space spanned by a subset of QITE states $|\Psi[s]\rangle$.

\section{Simulation Results and Experimental Realization}
\label{sec:sim_res}
\begin{figure}[ht!]
    \centering
    \includegraphics[scale=1.0]{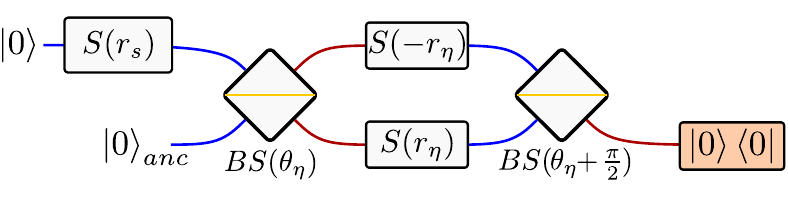}
    \caption{The quantum circuit for the implementation of QITE steps starting from initial state $|\Omega_0\rangle$ (Eq.\ \eqref{eq:Omega0}). }
    \label{fig:squeezer_circ}
\end{figure}
\begin{figure}[ht!]
    \centering
    \includegraphics[scale=1.5]{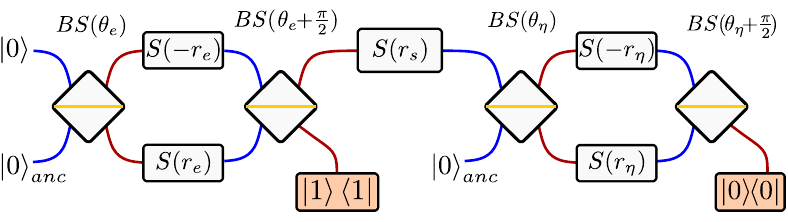}
    \caption{The quantum circuit for the implementation of QITE steps starting from an odd initial state $|\Omega(k)\rangle$ (Eq.\ \eqref{eq:33}).} 
    \label{fig:squeezer_circ_excited}
\end{figure}
In this section, we present our simulator results for the calculation of ground and excited state energies using our CV QITE algorithm. The quantum circuits depicted in Figures~\ref{fig:squeezer_circ} and~\ref{fig:squeezer_circ_excited} create approximations to the ground and first-excited states, respectively, as outlined above. {\color{black}{Even though these figures are for $L=1$-point case the generalization to $L$-point case is straightforward.}} These circuits can be easily simulated with Xanadu's Strawberry Fields photonic quantum computer simulator where the required states, Gaussian gates, and measurement tools are readily available. They can also be realized experimentally with existing technology.

\begin{figure}[ht!]
    \centering
    \includegraphics[scale=0.5]{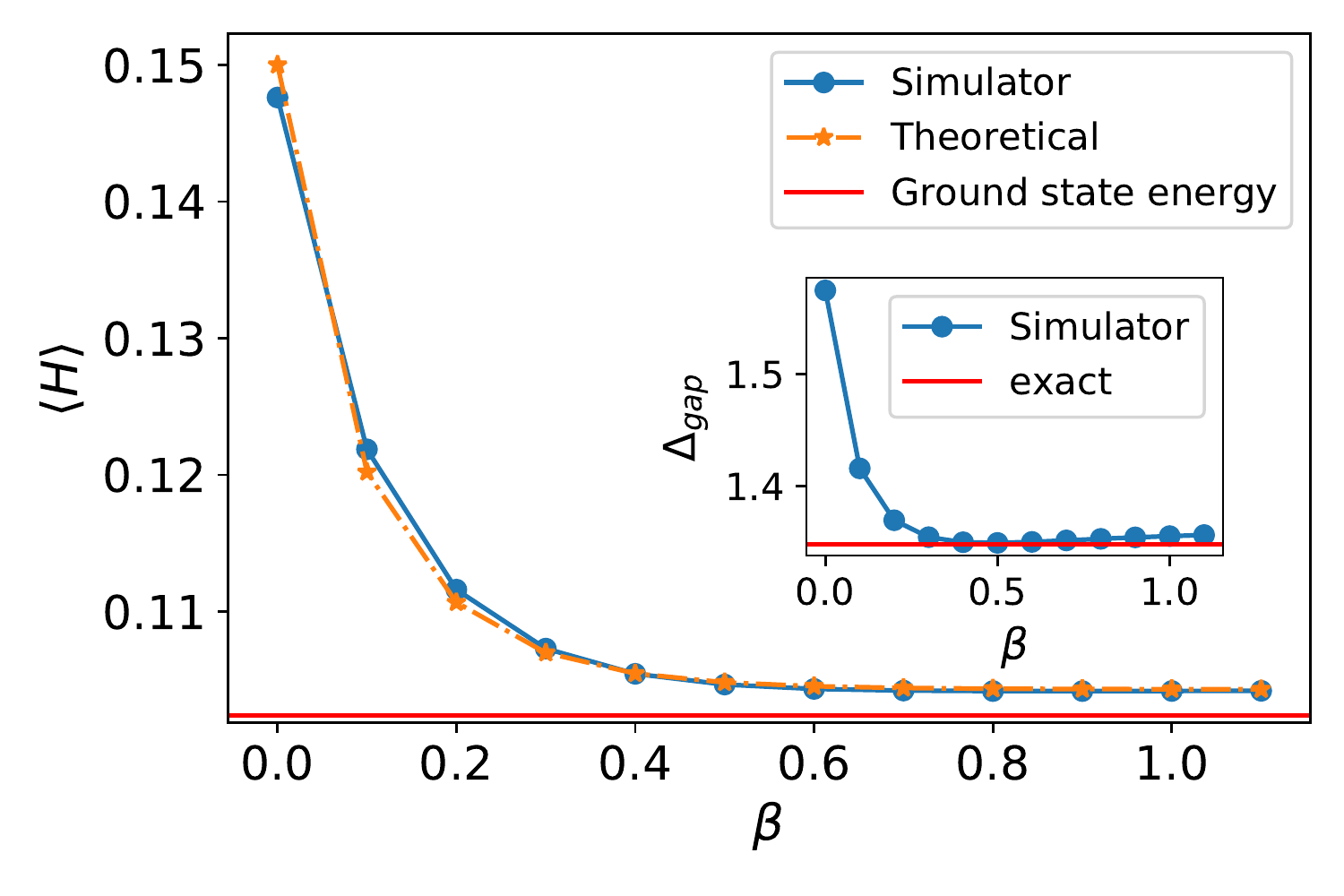}
    \caption{Energy expectation values as a function of imaginary time obtained using Xanadu's Strawberry Fields simulator and compared with theoretical values for the single lattice point toy model ($L=1$). Convergence to the ground state energy for $\lambda =4.8$ is observed with cutoff dimension $n_{\text{cutoff}} =10$ and $CX$ gate parameter $\eta=0.1$.} 
    \label{fig:onepointground}
\end{figure}

The energy expectation value obtained at each QITE step is plotted in Fig.~\ref{fig:onepointground} as a function of imaginary time for the single-point lattice, $L=1$, a toy model. We compare the values obtained from the simulator to exact analytic calculations for strength of interaction  $\lambda =4.8$, cutoff for the Hilbert space dimension in the simulator $n_{\text{cutoff}} =20$, and $CX$ gate parameter $\eta=0.1$. Starting the CV QITE algorithm with initial state $|\Psi[0]\rangle=|\Omega_0\rangle$, the energy expectation values are seen to converge to the ground state energy. Due to the cutoff in the Hilbert space dimension needed due to the limitations of the simulator, there is a 1.75\% error in the ground state energy. 

\begin{figure}[ht!]
    \centering
    \includegraphics[scale=1]{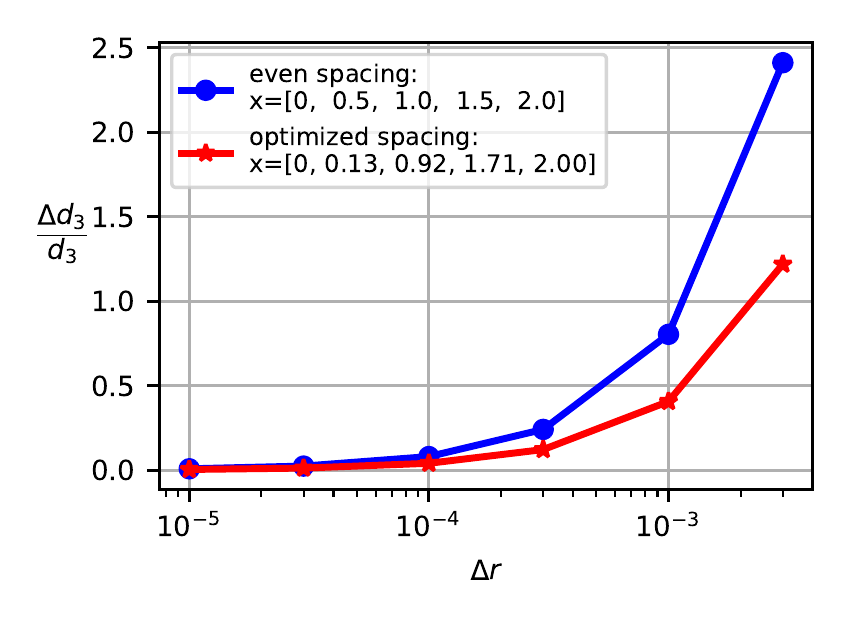}
    \caption{Relative uncertainty in the value of the third derivative in terms of the imprecision $\Delta r$ on the squeezing parameter in the $CX$ gate (Eq.\ \eqref{eq:23}).} 
    \label{fig:Delta_r}
\end{figure}

As discussed above, by choosing one of the odd wavefunctions $|\Omega(k)\rangle$ (Eq.\ \eqref{eq:33}) as initial state, the QITE algorithm approximates excited states of the system. The quantum algorithm in this case relies on the experimental setup depicted in Fig.~\ref{fig:squeezer_circ_excited} which can also be realized experimentally with existing technology. 
\begin{figure}[ht!]
    \centering
    \includegraphics[scale=0.36]{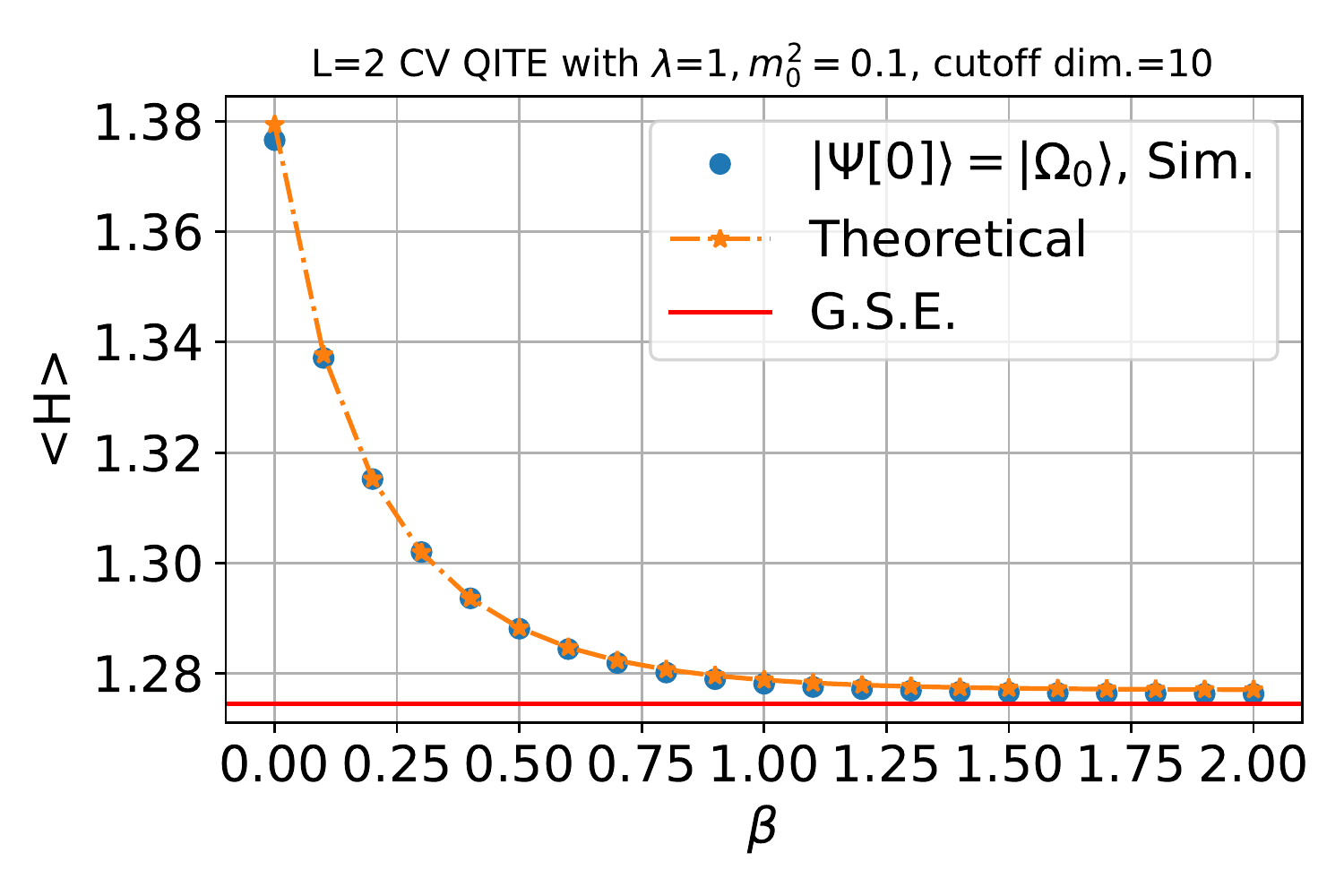}
    \includegraphics[scale=0.36]{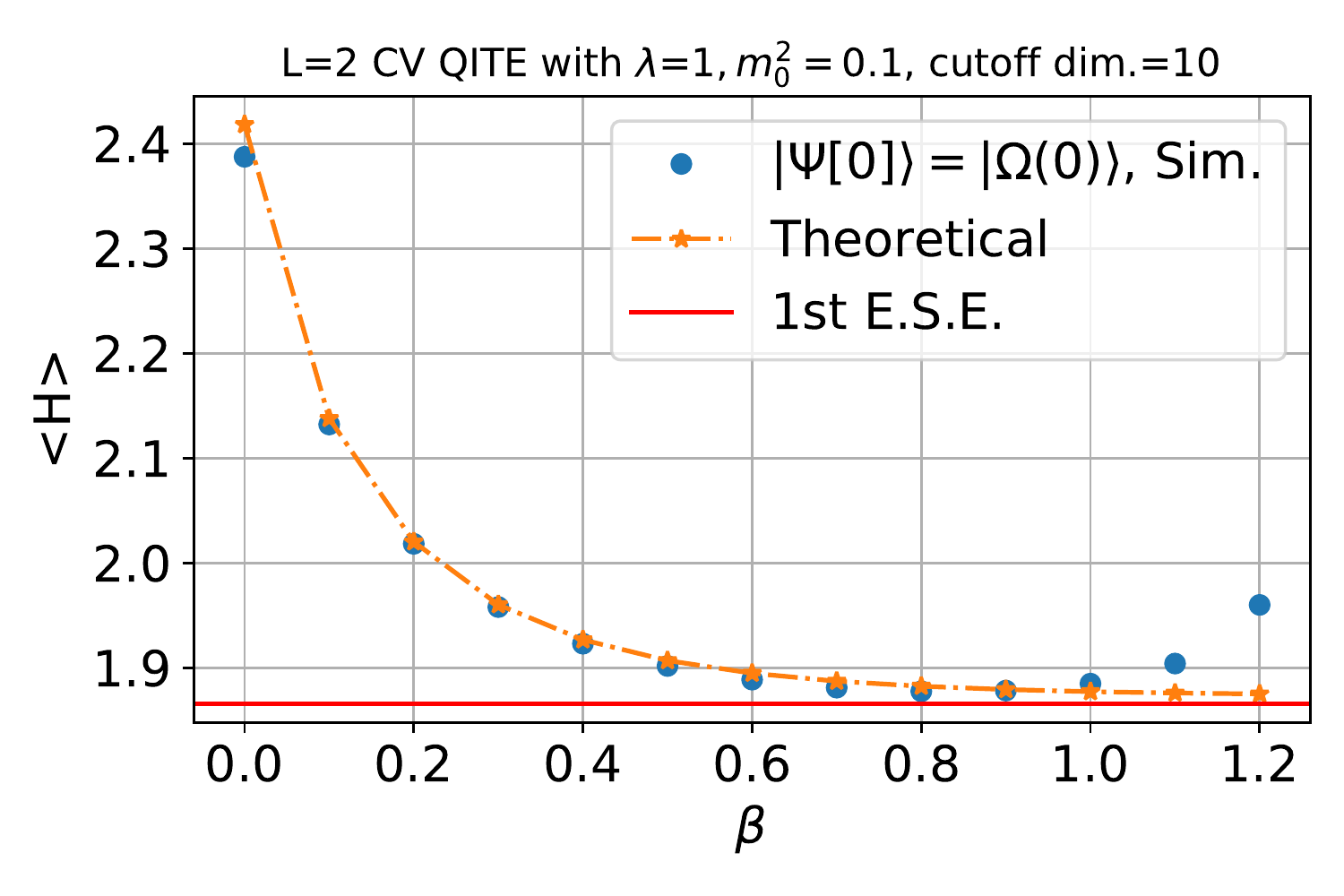}
    \includegraphics[scale=0.36]{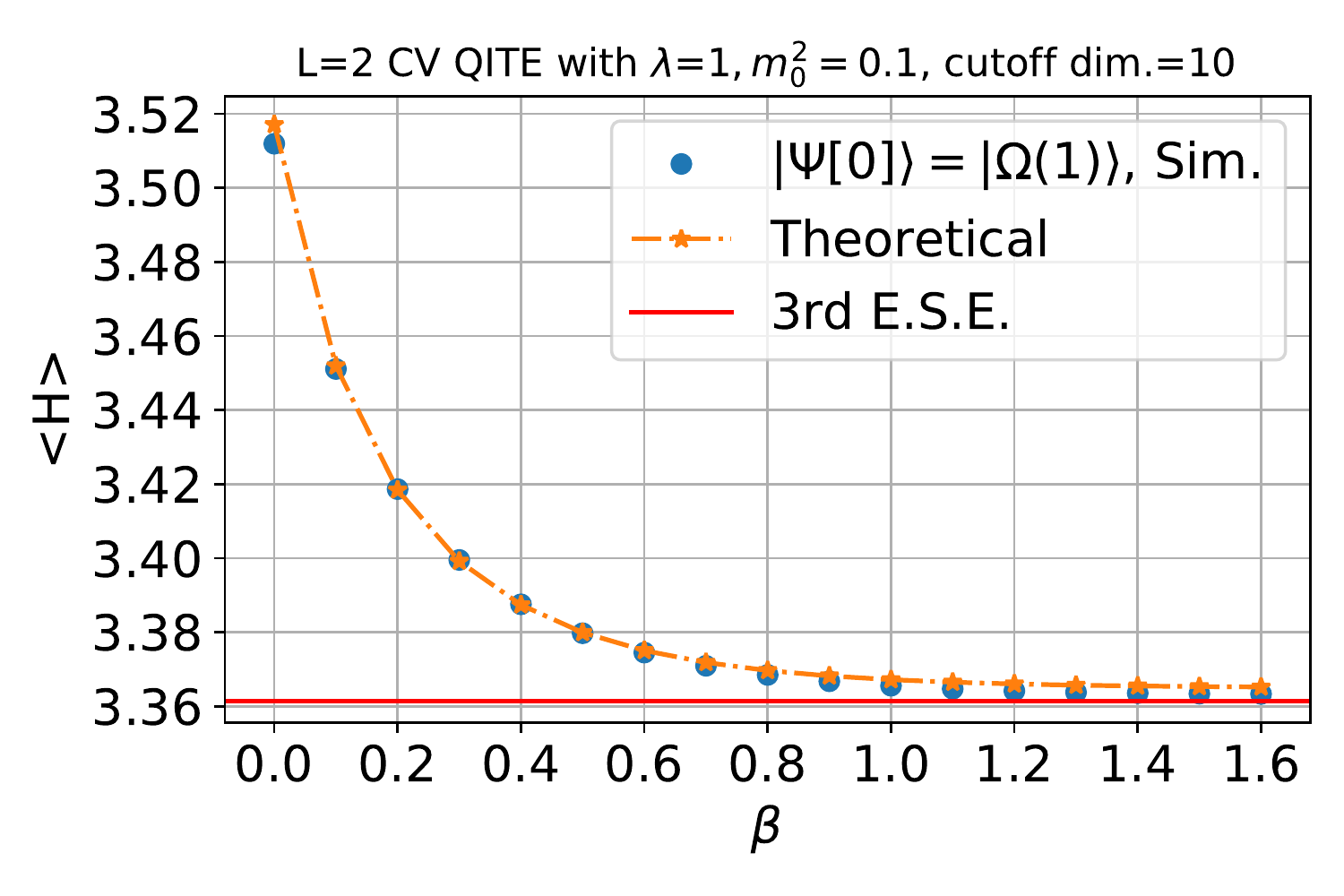}
    \caption{Energy expectation values as a function of imaginary time obtained with Xanadu's Strawberry Fields photonic simulator and compared with analytic values for $L=2$ lattice points. The parameters used are $\lambda =1$, $m_0^2=0.1$. We used cutoff dimension of Hilbert space in the simulator $n_{\text{cutoff}} = 10$ for convergence to the ground-state, the first and third excited-state energies.} 
    \label{fig:L2positivemsqenergies}
\end{figure}
In Fig.~\ref{fig:L2positivemsqenergies}, we demonstrate the convergence of the energy expectation values to the ground, first excited, and third excited state energies, starting with initial states $|\Omega_0\rangle$, $|\Omega(0)\rangle$, and $|\Omega(1)\rangle$, respectively, for the case of two lattice points ($L=2$). We used Xanadu's Strawberry Fields photonic simulator to obtain the simulator results. The parameters used were $m^2=0.1$, $\lambda =1$, and $\eta=0.1$. We ran the CV quantum circuit in Fig.~\ref{fig:squeezer_circ} at each QITE step for initial state $|\Omega_0\rangle$, and the one in Fig.\ \ref{fig:squeezer_circ_excited} for initial states $|\Omega(k)\rangle$ ($k=0,1$). The calculation of the various terms contributing to the parameters $\gamma_s(k)$ (Eq.\ \eqref{gamma_sq}) require additional measurements which were done with the aid of quantum circuits depicted in Fig.~\ref{fig:squeezer_circ_excited}. {\color{black}{In these calculations we only used the contributions coming from $\gamma_s(0)$ since it is the most contributing to the convergence of the energy expectation value. For this reason, the generalization of the quantum circuit to $L=2$-point case is just repetition of circuit in Fig.~\ref{fig:squeezer_circ} and Fig.~\ref{fig:squeezer_circ_excited} except an initial squeezer in the second qumode.}} We calculated the derivatives needed (Eq.\ \eqref{eq:q2nk}) using the differential quadrature method~\cite{Bert1996}. 
We used equal spacing of 0.1 between parameters in the derivative calculation in the $CX$ gate.
It should be noted that the calculation of high-order derivatives requires a certain precision in the parameters of the $CX$ gate (Eq.\ \eqref{eq:23}), the squeezing parameter $r$ and the beam splitter angle $\theta$, as well the measured probability $P_0(k)$ (Eq.\ \eqref{eq:30}). In Fig.~\ref{fig:Delta_r}, we show the relative uncertainty $ \frac{\Delta d_3}{d_3}$ in the value of the third derivative $d_3 = \frac{d^3 P_0}{d(\eta^2)^3}$ in terms of the uncertainty on the squeezing parameter $r$, for sample spacing $\eta^2_{i+1}-\eta^2_{i}$ values. Although we observe a sensitivity in the third derivative, we still achieve low relative uncertainty with reasonable sample spacing which can be further reduced by selecting appropriate, unevenly spaced samples.

\begin{figure}[ht!]
    \centering
    \includegraphics[scale=0.5]{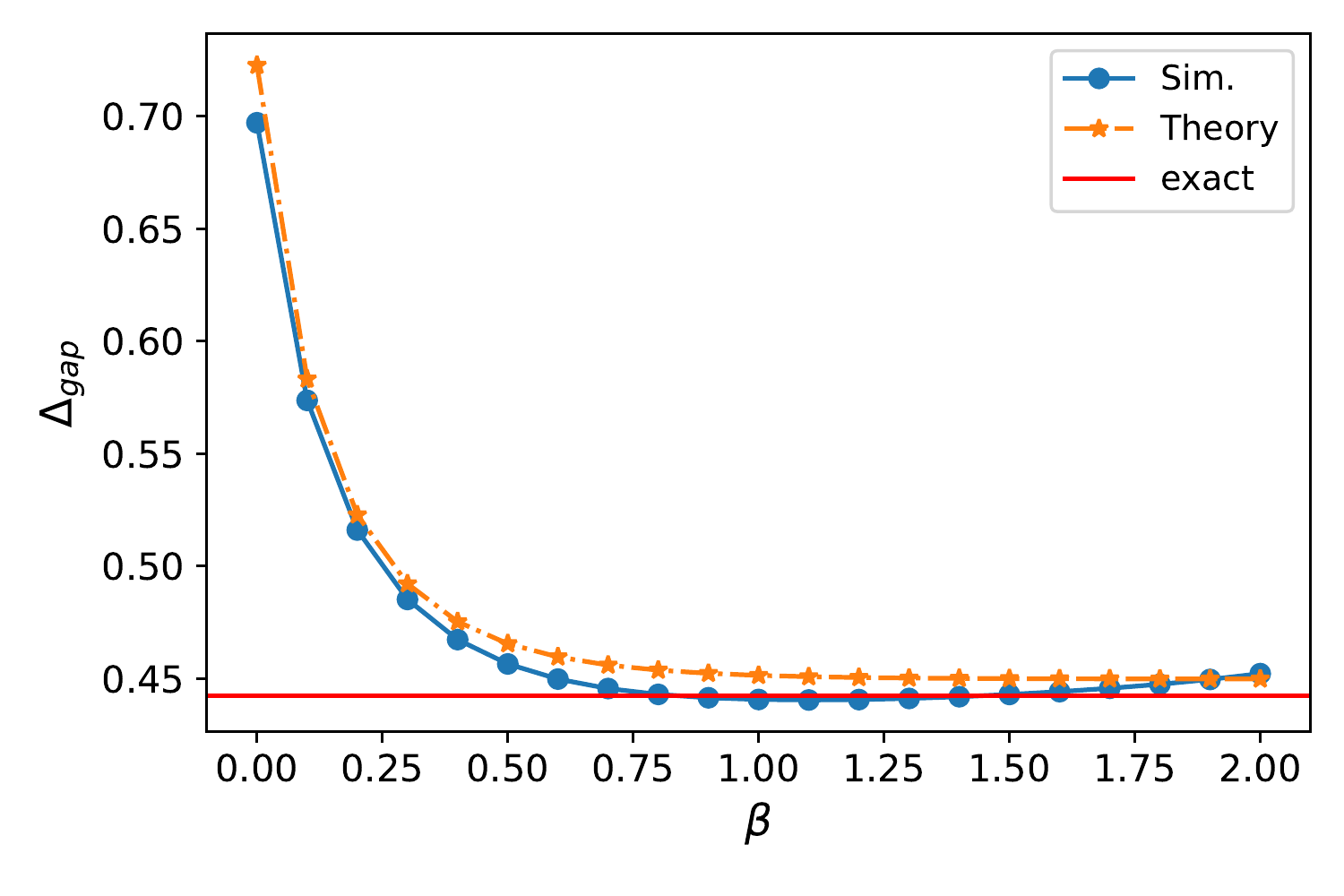}
    \caption{Mass gap values as a function of imaginary time obtained with Xanadu's Strawberry Fields photonic simulator and compared with analytic values for $L=2$ lattice points. The parameters used are $\lambda =1$, $m_0^2=-0.1$, $\delta m=0.2$.} 
    \label{fig:massgap}
\end{figure}

Fig.~\ref{fig:massgap} depicts the mass gap for the two-point lattice model ($L=2$) with parameters $m_0^2=-0.1$, $\delta m=0.2$, $\lambda = 1$. It converges rapidly to the expected value.

\section{Conclusion}
\label{sec:conclusion}

In this work, we developed a CV QITE algorithm to calculate the ground- and excited-state energies of a $\phi^4$ scalar QFT on a one-dimensional spatial lattice. Unlike with DV quantum computing, where a register of qubits is needed at each lattice point and the Hilbert space is truncated, only a single qumode at each lattice point was needed. The algorithm required ancilla qumodes for its implementation. The ancilla qumodes did not add a significant overhead as their number was comparable to the number of qumodes needed for the system. Our CV quantum algorithm avoided the use of non-Gaussian gates which are technologically challenging to implement. It only relied on Gaussian operators and measurements projecting onto photon number eigenstates. Therefore, it can be implemented with existing technology. Figures \ref{fig:squeezer_circ} and \ref{fig:squeezer_circ_excited} show the CV quantum circuits involved. 

We implemented our algorithm in simple cases of lattices with $L=1$ (toy model) and $L=2$ points using Xanadu's Strawberry Fields photonic quantum simulator. We observed convergence of energy expectation values to the ground- and excited-state energies depending on the choice of initial state, $|\Omega_0\rangle$ (Eq.\ \eqref{eq:Omega0}) and $|\Omega(k)\rangle$ (Eq.\ \eqref{eq:33}), respectively. Higher energy levels can also be reached by using the states derived in CV QITE as input to a CV version of the QLanczos algorithm.

Our results provide the basis for the development of CV quantum algorithms that rely on Gaussian operators and photonic measurements which can be implemented with existing technology. Our method can be extended to QFTs such as gauge theories that describe elementary particle interactions and the states derived with our CV QITE algorithm can be used for the calculation of various physical quantities of interest, such as scattering amplitudes. Work in this direction is in progress.

\acknowledgments

G.S.\ and E.M.\ are supported by ARO grant W911-NF-19-1-0397 and NSF grant OMA-1937008. K.Y.A.\ is supported by the U.S.\ Department of Energy, Office of Science, National Quantum Information Science Research Centers.

\end{document}